\documentclass[preprintnumbers, floatfix, twocolumn,
preprintnumbers, PRD, superscriptaddress,nofootinbib]{revtex4}
\usepackage{eurosym}
\usepackage{amsfonts}
\usepackage{amsmath}
\usepackage{amssymb,epsf}
\usepackage{latexsym}
\usepackage{graphicx,epsfig}
\usepackage{amssymb}
\usepackage{subfigure}
\usepackage{epstopdf}
\usepackage[colorlinks=true,citecolor=blue,linkcolor=blue,urlcolor=black]{hyperref}
\usepackage{color}

\graphicspath{{Images/}}

\renewcommand{\&}{\textup{\symbol{`\&}}}

\begin{document}

\title{Thermodynamic geometry and phase transition of spinning AdS black holes%
}
\author{Amin Dehyadegari}
\email{amindehyadegari@gmail.com}
\affiliation{Department of Physics, College of Sciences, Shiraz University, Shiraz 71454,
Iran}
\affiliation{Biruni Observatory, College of Sciences, Shiraz University, Shiraz 71454,
Iran}
\author{Ahmad Sheykhi}
\email{asheykhi@shirazu.ac.ir}
\affiliation{Department of Physics, College of Sciences, Shiraz University, Shiraz 71454,
Iran}
\affiliation{Biruni Observatory, College of Sciences, Shiraz University, Shiraz 71454,
Iran}

\begin{abstract}
Employing the thermodynamic geometry approach, we explore phase
transition of four dimensional spinning black holes in an anti-de
Sitter (AdS) spaces and found the following novel results. (i)
Contrary to the charged AdS black hole, thermodynamic curvature of
the spinning AdS black hole diverges at the critical point, without
needing normalization.(ii) There is a certain region with small
entropy in the space of parameters for which the thermodynamic
curvature is positive and the repulsive interaction dominates.
Such behavior exists even when the pressure is extremely large.
(iii) The dominant interactions in the microstructure of extremal
spinning AdS black holes are strongly repulsive, which is similar
to an ideal gas of fermions
at zero temperature. (iv) The maximum of thermodynamic curvature, $%
\left\vert R\right\vert $, is equal to $C_{{}_{P}}$ maximum values
for the Van der Waals fluid in the supercritical region. While for
the black hole, they are close to each other near the critical
point.
\end{abstract}

\maketitle


\section{Introduction \label{Intro}}

Thermodynamic fluctuation provides a unique frame for the geometrical
description of thermodynamical systems in equilibrium. Particular interest
goes to the covariant version, known as Ruppeiner geometry \cite{Rup1},
which consists of a metric that measures the probability of a fluctuation
between two thermodynamic equilibrium states. The Riemannian scalar
curvature, known as thermodynamic curvature, arises from such a metric is a
fundamental object in the Ruppeiner geometry which contains information
about inter-particles interaction. More specifically, a negative (positive)
sign of the thermodynamic curvature determines an attractive (repulsive)
interaction between particles. While zero value for the thermodynamic
curvature means there is no interaction between particles \cite%
{Rup2,Rup3,Rup4}. The absolute value of the thermodynamic curvature in the
asymptotic critical region is related to correlation length in fluids \cite%
{Rup3}.

Since the discovery of entropy and temperature of black holes \cite%
{beken,hawk}, it has been well established that one can regard black hole as
a thermodynamic system characterized by a set of thermodynamic variables.
During the past decades, various thermodynamic properties of black holes,
especially the phase transition and critical behavior, have been widely
studied in the literatures \cite{Davies,HawkingPage,VDW1,AAA}. In recent
years, considerable attentions have been arisen to investigate thermodynamic
phase transition of anti-de Sitter (AdS) black holes in an extended phase
space, where the first law of black hole thermodynamics is extended by
treating the cosmological constant as a thermodynamic variable \cite%
{Dolan1,Dolan2,PV,Dolan3,Dolan4,Dolan5,Hendi}. The investigations on
thermodynamic phase transition of black holes in the extended phase space
have disclosed some interesting phenomena, such as Van der Waals
liquid-vapor phase transition \cite{PV}, zeroth-order phase transition \cite%
{NAAA}, reentrant phase transition \cite{BIRPT,KerrRPT}, triple critical
point \cite{triplep}, superfluid like phase transition \cite{superfluidBH}\
and many others.

In the context of black hole thermodynamics, thermodynamic curvature in the
Ruppeiner geometry provides a powerful tool to explore microscopic behavior
of black holes. The obtained results can also be compared with accessible
experimental systems. Thermodynamic curvature has been investigated for
various types of black holes ( see e.g. \cite%
{Rup5,Rup6,Rup7,comment,Kord,MicroscopicRPT} and references
therein). It has been disclosed that thermodynamic curvature does
not diverge at the critical point, contrary to the fluid systems.
Recently, two new normalized thermodynamic curvature for a charged
AdS black hole have been proposed, which diverge at the critical
point of phase transition \cite{Wei1,Wei2,AAWei}. These
thermodynamic curvature are constructed via the heat capacity at
constant volume \cite{Wei1,Wei2} and adiabatic compressibility
\cite{AAWei} and have the same behavior for the large black hole.
In \cite{AAWei} it was shown that the normalized thermodynamic
curvature diverges to positive infinity for the extremal black
holes. More recently, the behavior of these two normalized
thermodynamic curvature was studied for several different black
holes \cite{Xu,Bhami1,Bhami2,Bhami3,Kumara1,Kumara2}.

In this paper, we explore thermodynamic phase structure of four-dimensional
rotating AdS black hole. We consider an extended phase space in the pressure
($P$) and entropy ($S$) plane, in which the small-like and large-like black
holes are separated by the maximum of the specific heat at constant pressure
in the supercritical region. Besides, we provide simple analytical
expressions for critical quantities. From the thermodynamic fluctuation
metric in the entropy representation, we obtain a Ruppeiner line element of
rotating-AdS black holes in the pressure-entropy coordinates, where it is
also valid for the ordinary thermodynamic systems, such as the simple Van
der Waals fluid. Then, by using the thermodynamic curvature, we explore the
microscopic properties of the system and compare it with the one of the Van
der Waals fluid system. In particular, we investigate the behavior of the
maximum of the specific heat at constant pressure and minimum of
thermodynamic curvature for these systems in the supercritical region. We
find that, for both cases, the thermodynamic curvature diverges at the
critical point and it goes to positive infinity for the extremal black
holes. Finally, the critical behavior of thermodynamic curvature for the
characteristic curves is studied and their critical exponents are calculated.

The rest of the paper is organized as follows. In Sec. \ref{TPH},we first
give a brief review on thermodynamics of four-dimensional rotating AdS black
hole in the extended phase space and then determine thermodynamic phase
structure in the $P$-$S$ plane. Next, we obtain the Ruppeiner metric in ($P$-%
$S$) coordinates, and using this, we study in details the microscopic
properties of the black hole and Van der Waals system in Sec. \ref{TC}.
Section \ref{CB} is devoted to investigating the thermodynamic curvature
near the critical region. In Sec. \ref{SD}, we present our summary and
discussion. In Appendix we calculate the thermodynamic curvature of Van der
Waals system using the Ruppeiner metric in ($P$-$S$) coordinates.

\section{Thermodynamic phase structure\label{TPH}}

\begin{figure}[t]
\epsfxsize=8.5cm \centerline{\epsffile{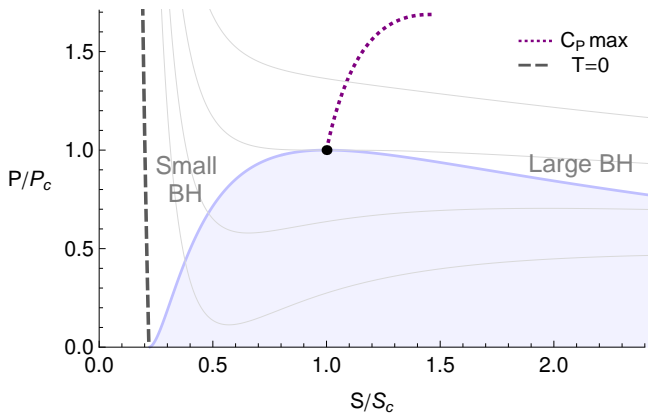}}
\caption{Phase structure of the rotating AdS black hole in $P$-$S$ plane.
The coexistence region of the first order phase transition between the large
and small black holes is identified. The gray dashed and purple dotted
curves correspond to the extremal black hole ($T=0$) and the local maximum
of $C_{{}_{P}}$, respectively; a black spot represents the critical point.
The several isothermal lines are denoted by the thin curves where, from top
to bottom, the temperature is decreased. The region on the left of the gray
dashed curve is excluded since it leads to negative temperature.}
\label{fig1}
\end{figure}

Let us begin with a brief review of the thermodynamics of single spinning
AdS black holes in four dimensions, based on Refs \cite{Cognola,Dolan2}. The
mass of the Kerr-AdS black hole with the pressure ($P$) is \cite{Dolan2}
\begin{equation}
M(S,P,J)=\frac{1}{2}\sqrt{\frac{(1+8SP/3)[4\pi ^{2}J^{2}+S^{2}(1+8SP/3)]}{%
\pi S}},  \label{Act1}
\end{equation}%
where $S$ and $J$ are the entropy and angular momentum, respectively. By
identifying the black hole mass as the enthalpy, the first law of
thermodynamics reads%
\begin{equation}
dM=TdS+\Omega dJ+VdP,
\end{equation}%
where $T$ is the Hawking temperature, $\Omega $ the angular velocity, $V$
the thermodynamic volume, which are given by
\begin{eqnarray}
\Omega &=&\frac{\pi J}{SM}(1+8SP/3), \\
V &=&\frac{2}{3\pi M}\left( S^{2}[1+8SP/3]+2\pi ^{2}J^{2}\right) , \\
T &=&\frac{1}{8\pi M}\left([1+8SP/3](1+8SP)-4\pi ^{2}J^{2}/S^{2}\right) .
\label{Tem}
\end{eqnarray}%
The internal energy $U$ is obtained from $M$ via the Legendre
transformation, $U=M-PV$, and it is given by
\begin{align}
&U (S,V,J)=\left(\frac{\pi }{S}\right)^{3}\Bigg{\{}\left(\frac{3V}{4\pi }%
\right)\left(\frac{S^{2}}{2\pi ^{2}}+J^{2}\right)  \notag \\
&-J^{2}\sqrt{\left(\frac{3V}{4\pi }\right)^{2}-\left(\frac{S}{\pi }%
\right)^{3}}\Bigg{\}}.
\end{align}%
In this representation, the first law of the black hole thermodynamics is
written as%
\begin{equation}
dU=TdS+\Omega dJ-PdV.  \label{FL}
\end{equation}
Now, we turn to study the critical behavior of the rotating-AdS black hole
by investigating the specific heat at constant pressure
\begin{equation}
C_{{}_{P}}=T\left. \frac{\partial S}{\partial T}\right\vert _{P},
\end{equation}%
where we have also fixed $J$. For constant $J$ and $P=P_{c}$, the value of
critical point can be determined by an inflection point%
\begin{equation}
\frac{\partial T}{\partial S}\Big|_{P_{c}}=0,\quad \quad \quad \frac{%
\partial ^{2}T}{\partial S^{2}}\Big|_{P_{c}}=0.  \label{cpoint}
\end{equation}%
Using the temperature formula in Eq.(\ref{Tem}), the critical quantities are
obtained analytically as
\begin{eqnarray}
S_{c} &=&\frac{24\pi J}{(73+6\sqrt{87})^{1/3}+(73-6\sqrt{87})^{1/3}-5}%
\approx 28.719J,  \notag \\
P_{c} &=&\frac{\left[ (73+6\sqrt{87})^{1/3}+(73-6\sqrt{87})^{1/3}-5\right]
^{2}}{768\pi J}\approx 0.003/J,  \notag \\
T_{c}^{2} &=&\frac{(6137+768\sqrt{87})^{1/3}-\frac{239}{(6137+768\sqrt{87}%
)^{1/3}}-7}{384\pi ^{2}J}\approx 0.002/J.  \notag \\
&&
\end{eqnarray}%
These quantities are the same, numerically, as ones found in Ref. \cite%
{Sha2016}. Here we present their analytical expressions for the
first time in a compact form. For $P>P_{c}$, the specific heat at
constant pressure is positive, i.e., black hole is
thermodynamically stable. However, below $P_{c}$, there exists a
certain range of quantities, for which the specific heat at
constant pressure is negative ($C_{{}_{P}}<0$). This corresponds
to a thermodynamic instability of the black hole which is remedied
by the Maxwell equal area construction, $\oint VdP=0$, indicating
a first order phase transition between small and large black
holes. The region of the first order phase transition, which is
obtained from the the Maxwell construction, is identified in the
$P$-$S$ plane in Fig. \ref{fig1}. The small and large black hole
phases are located at the left and right of the shaded region,
respectively. In Fig. \ref{fig1}, the extremal black hole curve
(corresponding to zero temperature) is denoted by the gray dashed
line and the critical point is indicated by a black solid circle.
The left region of the gray dashed curve is physically excluded
because the temperature becomes negative.

For the supercritical region, which is at higher pressures and entropies
than the critical point, we illustrate the local maximum of the specific
heat at constant pressure ($C_{{}_{P}}$) in Fig. \ref{fig1} by the purple
dotted line. The local maximum of $C_{{}_{P}}$ commences from $(\widetilde{P}%
,\widetilde{S})\approx (1.69,1.45)$ and terminates at the critical point,
where it goes to infinity and $\widetilde{P}=P/P_{c}$ and $\widetilde{S}%
=S/S_{c}$ are the reduced pressure and entropy, respectively. This curve can
be viewed as an extension to the coexistence line, which divides the
supercritical region into two phases \cite{PRB,Rup0}. Here, the small-like
and large-like black holes are separated by the local maximum of $C_{{}_{P}}$
in the supercritical region beyond the critical point.

\section{Thermodynamic curvature\label{TC}}

To set up a thermodynamic Riemannian geometry, we consider the rotating AdS
black hole in the canonical (fixed $J$) ensemble of extended phase space so
that its thermodynamic state is specified by the internal energy $U$ and
volume $V$. The line element of the geometry, which characterizes the
distance between thermodynamic states, is given by \cite{Rup1}
\begin{equation}
dl^{2}=-\frac{\partial ^{2}S}{\partial x^{\mu }\partial x^{\nu }}dx^{\mu
}dx^{\nu },  \label{metric}
\end{equation}%
where $S$ is entropy and $x^{\mu }=\left( U,V\right) $. Using the first law
for rotating AdS black hole Eq.(\ref{FL}) and the Maxwell relation, one can
express the line element Eq.(\ref{metric}) as follows\footnote{%
Although this line element is derived for the rotating AdS black hole, it
remains valid for an ordinary thermodynamic system \cite{AASH}.}
\begin{equation}
dl^{2}=\frac{1}{T}\left( \frac{\partial T}{\partial S}\right) _{P}dS^{2}-%
\frac{1}{T}\left( \frac{\partial V}{\partial P}\right) _{S}dP^{2}.
\label{Rupm}
\end{equation}%
By computing the Riemannian curvature scalar, $R$, (thermodynamic curvature)
from the metric, one can get some information about the interparticle
interaction in the thermodynamic system. In particular, the positive
(negative) sign of the thermodynamic curvature specifies that the dominant
interaction is repulsive (attractive) \cite{Rup2,Rup3,Rup4}. On the other
hand, $R=0$ shows there is no interaction in the system \cite{Rup8}. In what
follows, we examine the behavior of the thermodynamic curvature for the
rotating AdS black hole and the Van der Waals fluid.

\begin{figure}[t]
\epsfxsize=8.5cm \centerline{\epsffile{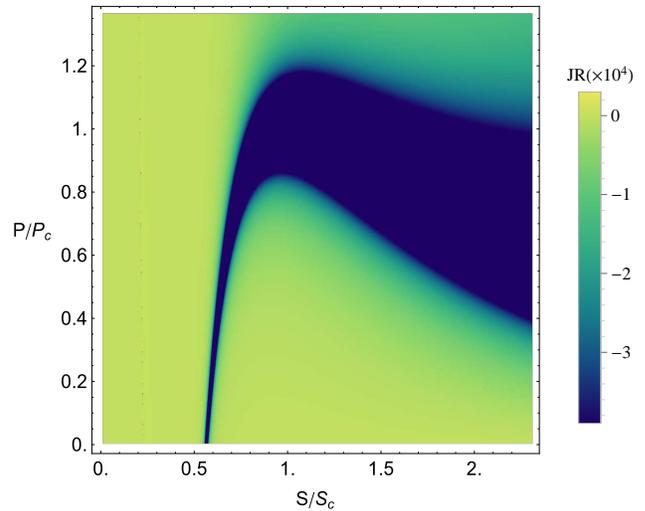}}
\caption{Bahavior of the thermodynamic curvature $R$ for the rotating AdS
black hole.}
\label{fig2}
\end{figure}

For the four-dimensional rotating AdS black hole, the thermodynamic
curvature is readily calculated as
\begin{equation}
R=\frac{\mathcal{B}(\widetilde{S},\widetilde{P})}{J\widetilde{T}[(\partial
\widetilde{T}/\partial \widetilde{S})_{\widetilde{P}}]^{2}},  \label{TCBH}
\end{equation}%
where $\mathcal{B}(\widetilde{S},\widetilde{P})$ is a complicated function
of the reduced pressure ($\widetilde{P}$) and entropy ($\widetilde{S}$) and $%
\widetilde{T}=T/T_{c}$ is the reduced temperature. Note that $R$ is
proportional to the inverse of angular momentum in the reduced parameter
space. The behaviour of $R$ is depicted in Fig. \ref{fig2} as a function of $%
P/P_{c}$ and $S/S_{c}$. One can see from Fig. \ref{fig2} that $R$ is
positive in some region of the parameter space. From Eq.(\ref{TCBH}), $R$
diverges on $\widetilde{T}=0$ and $(\partial \widetilde{T}/\partial
\widetilde{S})_{\widetilde{P}}=0$ corresponding to the extremal black holes
and diverging specific heat at constant pressure, respectively.
\begin{figure}[t]
\epsfxsize=8.5cm \centerline{\epsffile{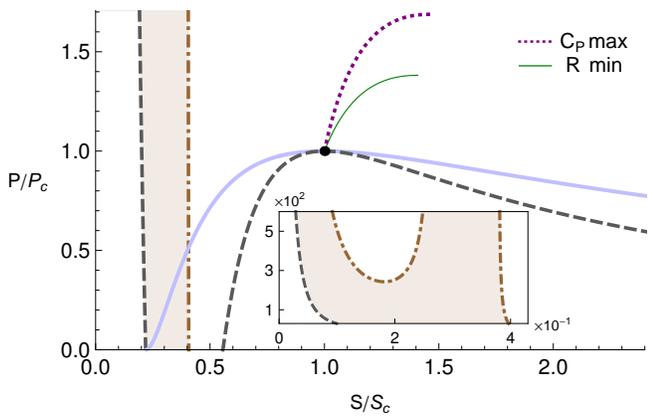}}
\caption{Coexistence curve (light blue solid line), diverging curve (gray
dashed line) and vanishing curve (brown dotted line) of $R$ for the rotating
AdS black hole. The region on the left of the gray dashed curve is excluded
since it leads to negative temperature. The thin green and purple dotted
curves correspond to the local minimum of $R$ and local maximum of $%
C_{{}_{P}}$, respectively; a black spot represents the critical point. In
the shaded region, $R$ is positive, otherwise $R<0$. The inset shows a
negative region in the shaded region for $\widetilde{P}\gtrsim 242.78$.}
\label{fig3}
\end{figure}

In order to examine the thermodynamic curvature more closely, we plot in
Fig. \ref{fig3} the vanishing (brown dotted line) and diverging (gray dashed
line) curves of $R$ as well as the transition curve (light blue solid line)
of small and large black holes and local maximum of $C_{{}_{P}}$ (purple
dotted line), which were shown already in Fig. \ref{fig1}. In Fig. \ref{fig3}%
, the shaded regions represent positive values of $R$, where the dominant
interaction is repulsive. In contrast, $R$ is negative everywhere outside
the shaded regions, indicating the dominant attractive interaction.
Remarkably, the transition and diverging curves coincide at the critical
point which is highlighted by a black spot. This situation also occurs for
ordinary thermodynamic systems \cite{Rup6}. The white area to the left of
the gray dashed line on the left side of the figure is excluded because of a
negative temperature. One can see from Fig. \ref{fig3}\ that the associated $%
R$ for the large black hole phase is negative. However, for the small black
hole phase, there exists a certain region with positive $R$, which is also
present in the higher pressure regime. In this region, when approaching the
gray dashed curve from above, $R$ diverges to $+\infty $ and dominant
interaction becomes strongly repulsive. The inset in Fig. \ref{fig3} reveals
the existence of a region with negative $R$ in the shaded region when $%
\widetilde{P}$ is greater than $\approx 242.78$. Moreover, in Fig. \ref{fig3}
we also display the local minimum of $R$ in the supercritical region by the
thin green line, which begins from $(\widetilde{P},\widetilde{S})\approx
(1.41,1.38)$ and ends at the critical point where $R$ goes to negative
infinity.

\begin{figure}[t]
\epsfxsize=8.5cm \centerline{\epsffile{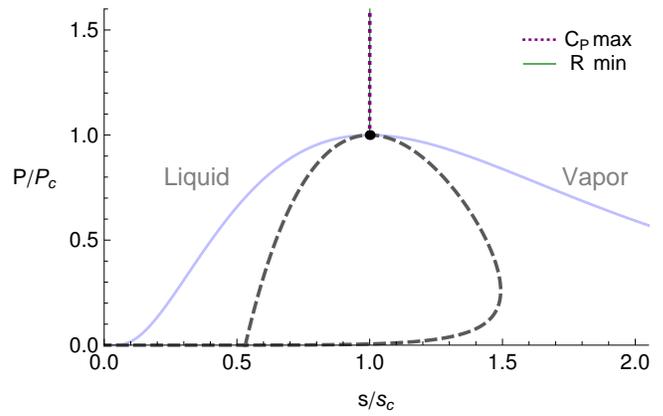}}
\caption{Coexistence curve (light blue solid line), diverging curve (gray
dashed line) of $R$ for the Van der Waals fluid. The thin green and purple
dotted curves correspond to the minimum of $R$ and maximum of $C_{{}_{P}}$,
respectively; a black spot represents the critical point. The thin green and
purple dotted curves match with each other. $R$ is negative everywhere in
this plane.}
\label{fig4}
\end{figure}

In Fig. \ref{fig4}, we depict the coexistence curve (light blue solid line)
of the Van der Waals vapor-liquid phase transition and maximum of $%
C_{{}_{P}} $ (purple dotted line) as well as the diverging (gray dashed
line) and minimum (thin green line) of $R$, where the expression of $R$ is
given in Appendix. According to Eq.(\ref{RVdW}) and Fig. \ref{fig4}, $R$ has
negative values everywhere, indicating the dominant attractive interaction
among the molecules. The coexistence and diverging curves coincide at the
critical point,which is marked by a black dot. Furthermore, as also seen in
Fig. \ref{fig4}, the maximum of $C_{{}_{P}}$ and minimum of $R$ curves match
each other in the supercritical region. For the region below the coexistence
curve, the Van der Waals model is inapplicable, so it is not considered
here.

\section{Critical properties\label{CB}}

\begin{figure}[t]
\epsfxsize=8.5cm \centerline{\epsffile{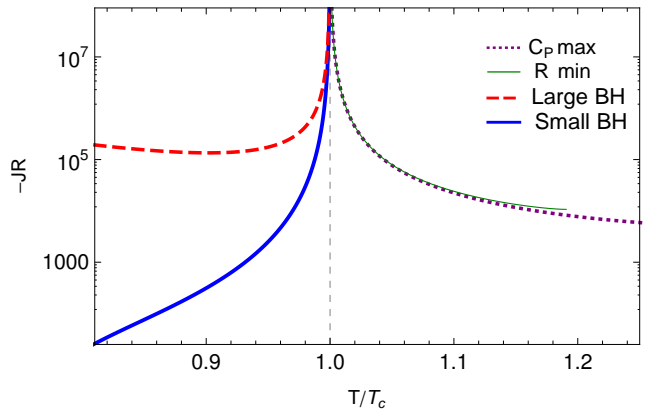}}
\caption{The thermodynamic curvature $R$ of characteristic curves for the
rotating AdS black hole. The purple dotted and thin green curves are close
together.}
\label{fig5}
\end{figure}

To further clarify the critical behavior of thermodynamic curvature for the
rotating AdS black hole and associated critical exponent, we investigate the
thermodynamic curvature of characteristic curves around the critical point.
To do so, in Fig. \ref{fig5}, we illustrate $R$ along its minimum and
maximum of $C_{{}_{P}}$ curves as well as along the transition curve for
small and large black holes in the neighborhood of the critical temperature.
As evident from the figure, the large black hole is at higher$\ \left\vert
R\right\vert $ than the small black hole and upon approaching the critical
point, $R$ in both phases diverges as
\begin{equation}
R\approx -\frac{41.2}{J}\left\vert t\right\vert ^{-2},
\end{equation}%
with a universal critical exponent of $2$, where $t=T/T_{c}-1$ is the
deviation from the critical temperature. In the supercritical regime, the
local minimum of $R$ and maximum of $C_{{}_{P}}$ curves are close together
in thermodynamic curvature and they diverge from above $T_{c}$ as%
\begin{equation}
R\approx -\frac{165.3}{J}t^{-2},
\end{equation}%
implying a critical exponent of $2$.

For the Van der Waals fluid, the thermodynamic curvature of the vapor and
liquid along the coexistence curve near the critical temperature has the
following form%
\begin{equation}
R=-\frac{1}{12}\left\vert t\right\vert ^{-2}.
\end{equation}%
Moreover, upon approaching the critical point from above along the minimum
of $R$ and maximum of $C_{{}_{P}}$ curves, $R$ diverges with the exponent $2$
as
\begin{equation}
R=-\frac{1}{3}t^{-2}.
\end{equation}

\section{Summary and discussion \label{SD}}

Thermodynamic geometry of black holes provides a powerful tool to explore
microscopic structure of these systems and disclose the nature of
interaction between their ingredient particles. In this paper, we have
presented simple exact analytical expressions for the critical quantities of
the Kerr-AdS black holes and constructed the phase diagram in the
pressure-entropy parameter space, where the small black hole and large black
hole phases are separated by a first order phase transition region below the
critical point. Based on the locus of the maxima of the specific heat at
constant pressure, we divided the supercritical region into small-like and
large-like black hole regions. Indeed, the line of maxima is used as the
Widom line, which is characterized by the maximum of the correlation length.
In addition, starting from the Ruppeiner geometry in an entropy
representation, we have derived the thermodynamic metric for the Kerr-AdS
black holes in the pressure-entropy coordinates that is also valid for any
ordinary thermodynamic system. We have explicitly shown that, contrary to
the charged AdS black hole \cite{AASH}, thermodynamic curvature of the Kerr
AdS black hole diverges at the critical point, without needing
normalization. Comparing to the simple Van der Waals fluid, which has
negative thermodynamic curvature everywhere, we have found that there is a
certain region for the spinning AdS black holes with small entropy in the
space of parameters for which the thermodynamic curvature is positive and
the repulsive interaction dominates. Such behavior exists even when the
pressure is extremely large. Another distinction is that the dominant
interactions in the microstructure of extremal Kerr AdS black holes are
strongly repulsive, which is similar to an ideal gas of fermions at zero
temperature \cite{Rup2}.

Taking into account the fact that the magnitude of the thermodynamic
curvature is related to the correlation length, we have used the locus of
the maximum of $\left\vert R\right\vert $ to characterize the Widom line. We
have found the maximum of $\left\vert R\right\vert $ is equal to $C_{{}_{P}}$
maximum values for the Van der waals fluid in the supercritical region.
While for the black hole, they are close to each other near the critical
point. Finally, we determined the critical behavior of thermodynamic
curvature of spinning AdS black hole and find out that governs by a universal
critical exponent of $2$, which is the same as the Van der Waals fluid.

It would be interesting to study reentrant phase transitions and universal
properties of higher dimensional rotating AdS black holes by employing the
thermodynamic Riemannian geometry based on the fluctuations of the entropy
and pressure.

\textbf{Note Added:} When this work was completed, we learned that
another article \cite{Weinew} had addressed the same issue where
it was shown that the thermodynamic curvature has different
behavior at small entropy. However, our results differ from
\cite{Weinew} in that we find a region within repulsive
interaction area in which the thermodynamic curvature has negative
values.



\appendix

\section{Van der Waals model \label{app1}}

In this Appendix, we calculate the thermodynamic curvature for Van der Waals
fluid in the $P$-$S$ plane. The specific Helmholtz free energy of the Van
der Waals, which contains two parameters ($a$,$b$) reflecting intermolecular
interaction and molecular size effects, is given by \cite{Landua}%
\begin{equation}
F=-\frac{a}{v}-T\left( \ln [v-b]+\frac{3}{2}\ln [T]+\ln [\zeta ]+1\right) ,
\label{FVdW}
\end{equation}%
where $\zeta =(m/2\pi )^{3/2}$ and $m$ is a mass of atom. Here, $T$ and $v$
are the temperature and specific volume, respectively. It is important to
note that $v>b$. Using Eq.(\ref{FVdW}), the pressure and entropy are
obtained as
\begin{equation}
\widetilde{P}=\frac{8\widetilde{T}}{3\widetilde{v}-1}-\frac{3}{\widetilde{v}%
^{2}},\quad \widetilde{s}=\frac{\widetilde{T}(3\widetilde{v}-1)^{2/3}}{%
2^{2/3}},
\end{equation}%
which is expressed in terms of the reduced thermodynamic variables
\begin{equation}
\widetilde{T}=\frac{T}{T_{c}},\quad \widetilde{P}=\frac{P}{P_{c}},\quad
\widetilde{v}=\frac{v}{v_{c}},\quad \widetilde{s}=\frac{s}{s_{c}},  \notag
\end{equation}%
where $s\equiv e^{(2S-5)/3}/\zeta ^{2/3}$ and $S$ is entropy. The critical
quantities are
\begin{equation}
P_{c}=\frac{a}{27b^{2}},\quad v_{c}=3b,\quad s_{c}=\frac{2^{11/3}a}{27b^{1/3}%
},\quad T_{c}=\frac{8a}{27b}.
\end{equation}%
Using the line element in ($P$-$S$) coordinates Eq.(\ref{Rupm}), the
thermodynamic curvature is obtained as
\begin{equation}
R=\frac{(3\widetilde{v}-1)^{8/3}[(3\widetilde{v}-1)^{8/3}-2^{11/3}\widetilde{%
s}\widetilde{v}^{3}]}{3[(3\widetilde{v}-1)^{8/3}-2^{8/3}\widetilde{s}%
\widetilde{v}^{3}]^{2}},  \label{RVdW}
\end{equation}%
which is independent of $a$ and $b$.

\end{document}